\documentclass[twocolumn, superscriptaddress, floatfix, showpacs, aps, prb, 10pt]{revtex4-2}
\pdfoutput=1
\usepackage{graphicx}
\usepackage{amsmath}
\usepackage[colorlinks=true, citecolor=blue, urlcolor=blue]{hyperref}
\usepackage{amssymb}
\usepackage{bm}
\usepackage{color}
\usepackage{array}
\usepackage{booktabs}
\usepackage{multirow}
\usepackage{braket}
\usepackage{siunitx,physics}

\newcommand{\pd}{{\phantom{\dag}}}

\begin{document}

\title{Topological properties of domain walls in antiferromagnetic topological insulators}

\author{Gabriele Naselli}
\affiliation{Institute for Theoretical Solid State Physics, IFW Dresden and W\"{u}rzburg-Dresden Cluster of Excellence ct.qmat, Helmholtzstr. 20, 01069 Dresden, Germany}
\affiliation{Dipartimento di Fisica “E. R. Caianiello”, Universitá di Salerno, IT-84084 Fisciano, Italy}

\author{Ion Cosma Fulga}
\affiliation{Institute for Theoretical Solid State Physics, IFW Dresden and W\"{u}rzburg-Dresden Cluster of Excellence ct.qmat, Helmholtzstr. 20, 01069 Dresden, Germany}

\begin{abstract}
Motivated by the study of stacking faults in weak topological insulators and the observation of magnetic domain walls in MnBi$_{2n}$Te$_{3n+1}$, we explore the topological properties of magnetic domain walls in antiferromagnetic topological insulators. 
We develop two tight-binding models for two different types of antiferromagnetic topological insulators: the first type obtained by adding antiferromagnetic order to a strong topological insulator, and another built from stacked Chern insulating layers with alternating Chern numbers.
Both systems are dual topological insulators, i.e. they are at the same time antiferromagnetic and crystalline topological insulators, but differ by the type of mirror symmetry protecting the crystalline phase: spinful versus spinless.
We show that in the spinful case the mirror Chern number is invariant under time reversal and that it changes sign in the spinless case.
This influences the properties of the two systems in the presence of a magnetic domain wall, which we model as an interface between two regions of opposite magnetization.
In the first type, the bulk of the magnetic domain wall is gapped but the defect will host chiral edge states when it ends on an external ferromagnetic surface. 
In the second, due to the change in the sign of the mirror Chern number, the magnetic domain wall is a two-dimensional embedded semimetal with 2D gapless states protected by mirror symmetry. 
Our results show that magnetic domain walls can be a source of non-trivial topology, allowing to generate and manipulate gapless states within the bulk and the ferromagnetic surfaces of antiferromagnetic topological insulators.
\end{abstract}

\maketitle

\section{Introduction}
\label{sec:intro}
In recent years there has been a growing interest in topological magnetic materials \cite{Bernevig2022, Wang2023}, such as ferromagnetic and antiferromagnetic topological insulators \cite{Chang2013, Checkelsky2014, Otrokov2019, gong2019experimental, Li2019_2, Eremeev2021, Varnava2022, Roychowdhury2023, Jiang2023, Liu2024} and semimetals \cite{Fang2012, Tang2016, Yang2017, Belopolski2019, Nie2019, Cano2019} and topological altermagnets \cite{Li2024, Ma_2024}.
One of the most well-known classes of topological magnetic materials are 
antiferromagnetic topological insulators (AFTIs) \cite{PhysRevB.81.245209}. AFTIs are three-dimensional (3D) topological insulators with antiferromagnetic order.
The first proposed type of three-dimensional topological insulators are weak and strong topological insulators (STIs) \cite{Fu2007}, which host surface states protected by time-reversal symmetry. 
Their presence is related via the bulk-boundary correspondence to a non-trivial $\mathbb{Z}_2$ topological invariant \cite{Fu2006, Moore2007, Fu2007}.

In AFTIs, the presence of antiferromagnetic order breaks time-reversal symmetry ($\mathcal{T}$) while preserving the combination $S=\mathcal{T} T_{1/2}$, where $T_{1/2}$ is a lattice translation by half of a unit cell \cite{PhysRevB.81.245209, Fang2013}. 
Similar to the time-reversal symmetric case, the bulk of the system has a non-trivial $\mathbb{Z}_2$ topological invariant, and gapless states protected by the new symmetry $S$ are present at some of the external surfaces \cite{PhysRevB.81.245209}. 
In these systems, there are two types of external surfaces: type A surfaces, where the $S$ symmetry is preserved and topological surface states occur, and type F surfaces, where the $S$ symmetry is broken and no topological states are present.

Lattice defects can negatively impact the topological features of different systems \cite{Fulga2014}, especially in AFTIs, weak topological insulators, and topological crystalline insulators (TCIs) \cite{Fu2011, Hsieh2012, Ando2015}, which rely on lattice symmetries for the protection of their topological states \cite{Zhang2018}.
Despite this, defects can also act as a source of non-trivial topology \cite{Teo2010, Ran2009, Teo2017, Jurii2012, Hughes2014, Teo2013, Benalcazar2014}, depending on their geometry and symmetries.
Recently, two-dimensional (2D) defects, such as stacking faults, have been shown to host non-trivial gapless electronic states in weak topological insulators \cite{Naselli2022a}. 
Given the similarities between the topological properties of weak topological insulators and AFTIs, we are interested in studying two-dimensional lattice defects in AFTIs. 

We consider two different types of AFTI, which we label as type-I and type-II.
The first (type-I) is made by adding antiferromagnetic order to an STI. A well-known AFTI in this class is MnBi$_{2n}$Te$_{3n+1}$ \cite{Otrokov2019, gong2019experimental}. 
This material is composed of a stack of MnBi$_2$Te$_4$ magnetic septuple layers with $n-1$ quintuple layers of Bi$_2$Te$_3$ between them. 
The magnetization of the Mn atoms orders antiferromagnetically in the stacking direction.
Bulk MnBi$_{2n}$Te$_{3n+1}$ is an AFTI with gapless states on type A surfaces and with gapped F surfaces \cite{Otrokov2019, gong2019experimental}. 
However, the material can exhibit different topological phases depending on the geometry, making it an important platform for the study of magnetic topological phases. 
For example, the topology changes in the quasi 2D limit, when a stack consisting of only a few layers is considered. 
When the system has an odd number of layers, it is a Chern insulator \cite{Otrokov2017, Otrokov2019_qhe, Li2019, Deng2020}. 
In the even layer case, the quasi-2D stack is not in a Chern insulating phase, but the so-called quantum layer-Hall effect arises in the presence of an external magnetic field \cite{Gao2021, Dai2022}.

Defects, such as magnetic domain walls that break the $S$ symmetry \cite{Lin2022} and step edges \cite{Xu2022},  have already been studied in this compound family. 
Interestingly, the latter host topological modes \cite{Xu2022}, due to the presence of a half-integer Chern number on the ferromagnetic surfaces of the material.
Magnetic domain walls that do not break the $S$ symmetry have been observed in MnBi$_2$Te$_4$ \cite{Sas2020, Wenbo2022} and their topological properties have been studied in stacks with an odd number of layers \cite{Liang2023} and at the intersection with the external ferromagnetic surfaces of the system \cite{Varnava2021, Rusinov2021}.  In these cases, the defect has been predicted to host chiral modes. 

The second type of AFTI that we study (type-II) is made up of a stack of weakly-coupled Chern insulating layers, with adjacent layers carrying opposite Chern numbers. 
This second type of topological phase does not appear in MnBi$_{2n}$Te$_{3n+1}$, but it could be realized in van der Waals materials, e.g., in stacks of graphene layers with opposite out-of-plane magnetization \cite{Das2019}. 
When a mirror symmetry is present, both types of AFTI are so-called dual topological insulators (DTIs) \cite{Facio2019}, which means that they are at the same time AFTIs carrying a non-trivial $\mathbb{Z}_2$ invariant and TCIs carrying a non-trivial mirror Chern number \cite{Fu2011, Hsieh2012, Ando2015, Das2019}. 

\begin{figure}
\centering
\includegraphics[width=0.75\linewidth]{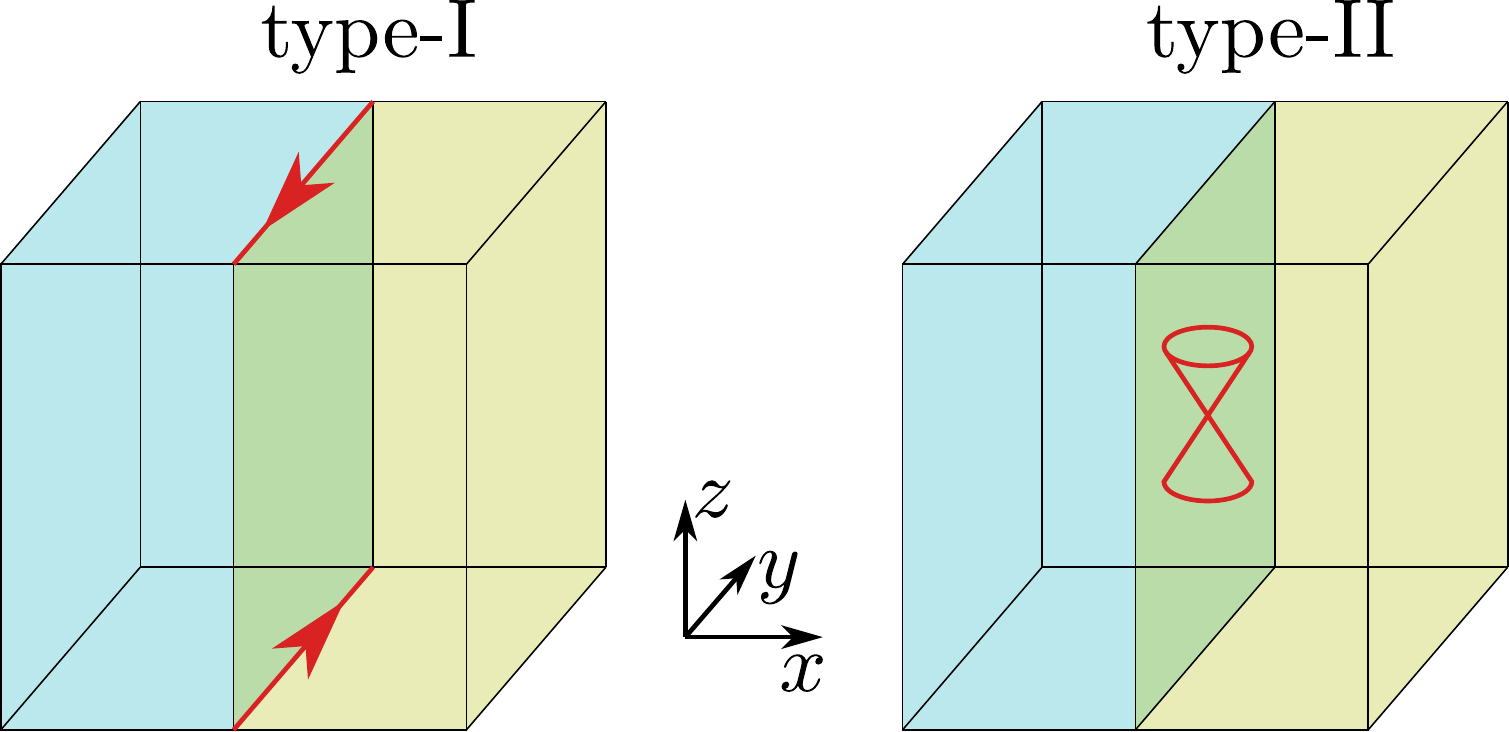}
\caption{
We consider magnetic domain walls (vertical green plane) formed between two regions with opposite magnetization (blue/yellow) in AFTIs. The domain wall is a 2D structure that lies in the bulk of the 3D AFTIs and separates regions of the systems with opposite magnetization. We refer to the intersection between the domain wall and the external surfaces of the system as boundaries of the domain wall, whereas we refer to the part of the domain wall that lies away from the external surfaces as bulk of the 2D structure.
In the type-I AFTI, the domain wall hosts chiral edge modes (red arrows), provided it ends on one of the ferromagnetic surfaces.
In the type-II AFTI, the domain wall is an embedded topological semimetal hosting Dirac cones protected by mirror symmetry, with gapless states localized in the bulk of the defect (green plane).
}
\label{fig:afti_domain_wall}
\end{figure}

In this work, we consider magnetic domain walls created by flipping the magnetization in part of the system.
Thus, the domain wall is formed as an interface between two systems related to each other by a time-reversal transformation.
Since magnetic domain walls are the only type of domain walls that we study in this work, the terms domain wall and magnetic domain wall are used interchangeably in the rest of the manuscript.
Our main findings are summarized in Fig.~\ref{fig:afti_domain_wall}.
We are interested in studying the topological properties of the bulk of the defect.
The domain wall is a 2D structure formed inside the 3D bulk of the material, shown in green in Fig.~\ref{fig:afti_domain_wall}. We refer to the intersection between the 2D defect and the external surfaces of the system as boundaries of the domain wall, whereas we refer to the part of the defect buried in the bulk of the AFTI as bulk of the domain wall.

The topological properties of the defect will depend on the effect of time reversal on the mirror Chern number of the system.
As we show, in type-I AFTIs time reversal does not change the mirror Chern number. 
In this case the bulk of the domain wall is gapped, but chiral edge states, shown in red in Fig.~\ref{fig:afti_domain_wall}, are present when the defect ends on an external F surface, similarly to Ref.~\cite{Varnava2021, Rusinov2021}.
In the type-II AFTI on the other hand, the mirror Chern number changes sign under time reversal. 
In this case, the domain wall hosts 2D gapless modes protected by mirror symmetry, shown as a Dirac cone in Fig.~\ref{fig:afti_domain_wall}, since it separates two regions with opposite mirror Chern number.

The rest of this work is organized as follows.
In Sec.~\ref{sec:model} and in Sec.~\ref{sec:DTI} we build tight-binding models for the two different types of AFTI. 
We show that each of the two systems has either a spinful or a spinless mirror symmetry associated with the crystalline topology. 
In Sec.~\ref{sec:trandmirrorchernnumber} we explore the effect of a time-reversal transformation on the mirror Chern number. 
We show that the mirror Chern number of TCIs with a spinful mirror symmetry is invariant under time reversal, while this topological invariant changes sign under time reversals for systems with spinless mirror symmetry. 
In Sec.~\ref{sec:domainwall} we confirm our analysis numerically, by showing that the domain walls in the toy model for the type-I AFTI are gapped, and that domain walls in type-II AFTIs host 2D gapless states which are protected by mirror symmetry.  
We then conclude in Sec.~\ref{sec:conclusion}.

\section{Antiferromagnetic topological insulator from strong topological phase}
\label{sec:model}

In this section, we build a tight-binding model for the type-I AFTI by adding a staggered magnetization to the Hamiltonian of a strong topological insulator, similar to Refs.~\cite{PhysRevB.81.245209, Naselli2022}.
We start with the Hamiltonian of an STI defined on a cubic lattice with lattice constant $a=1$ and two spinful orbitals per site, which we describe using two sets of Pauli matrices: $\bm{\sigma}$ for spin and $\bm{\tau}$ for the orbital \cite{Hosur2010}. 
The STI Hamiltonian is:
\begin{equation}
H_{\rm STI}(\vb{k})=B_0+\sum_{\mu=x,y,z} B^\pd_\mu e^{ik_\mu} + B^\dag_\mu e^{-ik_\mu},
\end{equation}
where $\vb{k}$ is the crystal momentum, $B_0=m\tau_z\sigma_0$ is an onsite term and $B_\mu=\frac{1}{2}\left(iv\tau_x\sigma_\mu+t\tau_z\sigma_0\right)$ are the hoppings in the $\mu=x, y, z$ directions. 
We double the unit cell in the $z$ direction such that $a_z \rightarrow 2a_z$ and $k_z \rightarrow k_z/2$.
In this way, the unit cell of the system contains two lattice sites stacked on top of each other.
By adding a staggered Zeeman term $B_M=M\tau_0\sigma_z$ we obtain the Hamiltonian of a type-I AFTI:
\begin{equation}
\label{eq:afti}
H_{\rm I}(k_x, k_y, k_z)= \begin{pmatrix}
H_0(k_x, k_y)+B_M & B^\pd_z+B_z^\dag e^{-ik_z}\\
B^\dag_z+B_z^\pd e^{ik_z} & H_0(k_x, k_y)-B_M
\end{pmatrix},
\end{equation}
where $H_0(k_x, k_y)$ is the layer Hamiltonian:
\begin{equation}
H_0(k_x, k_y)=B_0+\sum_{\mu=x,y} B^\pd_\mu e^{ik_\mu} + B^\dag_\mu e^{-ik_\mu}.
\end{equation}

The system is symmetric under a combination of time-reversal and translation symmetry $S(k_z)=\mathcal{T}T_{1/2}(k_z)$,  where the time-reversal and the translation operators take the form:
\begin{equation}
\mathcal{T}=i\tau_0\sigma_y\begin{pmatrix}
1 & 0 \\
0 & 1
\end{pmatrix}\mathcal{K} \quad T_{1/2}(k_z)=\tau_0\sigma_0\begin{pmatrix}
0 & 1 \\
e^{ik_z} & 0
\end{pmatrix},
\end{equation}
where the matrices written out in $2\times2$ form characterize the layer degree of freedom and ${\cal K}$ denotes complex conjugation. The symmetry constraint in momentum space is:
\begin{equation}
S(k_z)H_{\rm I}(k_x, k_y, k_z)S^{-1}(k_z)=H_{\rm I}(-k_x, -k_y, -k_z). 
\end{equation}
On the $k_z=0$ plane the symmetry constraint takes the form:
\begin{equation}
S(0)H_{\rm I}(k_x,k_y, 0)S^{-1}(0)=
H_{\rm I}(-k_x,-k_y, 0),
\end{equation}
with $S^2(0)=-1$.  
Setting, $t=1$, $m=0.5$, $v=2$, and $M=1$, we find that the two-dimensional Hamiltonian $H_{\rm I}(k_x,k_y, 0)$ carries a non-zero $\mathbb{Z}_2$ topological invariant $\nu=1$ \cite{PhysRevB.81.245209}. 
This $\mathbb{Z}_2$ invariant is analogous to the one used to classify two-dimensional systems with quantum spin-Hall effect \cite{Kane2005, Kane2005_2, Fu2006, Moore2007, Fu2007}, where  $S(k_z=0)$ plays the role of the time-reversal operator $\mathcal{T}$.

Since $\nu$ is non-zero, the system is an AFTI and its external surfaces host gapless topological states. 
We can distinguish two types of external surfaces in the system. 
The surfaces which preserve the antiferromagnetic order, type A, will host topological surface states. 
For example, the $(100)$ surfaces in our model host topological surface states associated with a Dirac cone at $k_z=0$ in the surface band structure. 
The surfaces with ferromagnetic order, type F, e.g. the top and bottom surfaces $(001)$ and $(00\bar{1})$, do not host gapless surface states. 
This is due to the fact that the ferromagnetic order locally breaks the $S$ symmetry. 
The F surfaces are not topologically trivial, but they carry a half-integer Chern number \cite{PhysRevB.81.245209}:
\begin{equation}
C_{T/B} =\pm\frac{1}{2},
\end{equation}
which enables chiral modes to form when a domain wall appears at the surface.

The system has also a spinful mirror symmetry with respect to the $k_z=0$ plane:
\begin{equation}
M_z(k_z)H_{\rm I}(k_x, k_y, k_z)M_z^{\dag
}(k_z)=H_{\rm I}(k_x, k_y, -k_z),
\end{equation}
where the symmetry operator $M_z(k_z)$ is defined as:
\begin{equation}
M_z(k_z)=-i\tau_z\sigma_z\begin{pmatrix}
1 & 0 \\
0 & e^{-ik_z}
\end{pmatrix}.
\end{equation}
The system is thus a dual topological insulator (DTI) \cite{Facio2019}, which has both a non-trivial $\mathbb{Z}_2$ invariant and a non-trivial mirror Chern number. 
To evaluate the mirror Chern number we block diagonalize the Hamiltonian on the mirror plane $k_z=0$, by rotating in the basis in which the mirror operator takes the form:
\begin{equation}
\begin{split}
\tilde{M}_z(k_z=0)=U  M_z(0)U^\dag =\begin{pmatrix}
i \mathbb{I}_{4\times4} & 0 \\
0 & -i\mathbb{I}_{4\times4}
\end{pmatrix}.
\end{split}
\end{equation}
In the new basis the Hamiltonian is:
\begin{equation}
\tilde{H}_{\rm I}(k_x, k_y, 0)=\begin{pmatrix}
H_{m_+}(k_x, k_y) & 0 \\
0 & H_{m_-}(k_x, k_y)
\end{pmatrix},
\end{equation}
where the blocks $H_{m_\pm}(k_x, k_y)$ are associated to opposite mirror eigenvalues $m_\pm=\pm i$. 
The two blocks have opposite Chern numbers, $C(H_{m_\pm})=\pm 1$, and the 3D Hamiltonian carries a mirror Chern number given by the difference \cite{Hsieh2012, Ando2015, Das2019}:
\begin{equation}
C_M=\frac{C(H_{m_+})-C(H_{m_-})}{2}=1.
\end{equation}
The dual topology does not change the physical properties of the clean system, apart from offering extra symmetry protection to its surface states. 
In this type of AFTI, the dual topology does not play a role in the protection of topological states in the presence of a domain wall either. 

This is due to the fact that the AFTI inherits the dual topology from the STI phase. 
In the time-reversal invariant case, for $M=0$, the system is an STI, so the (100) mirror-symmetric surface will host a surface Dirac cone at $k_z=0$. 
The states $\ket{\Psi}$ and $\mathcal{T}\ket{\Psi}$ that cross at the Dirac point are time-reversal conjugate of each other, but at the same time they are also eigenstates of mirror symmetry.
Since the operators $\mathcal{T}$ and $M_z(k_z)$ commute, states that belong to the same Kramers pair have to carry opposite mirror eigenvalue $m= \pm i$, i.e.
\begin{equation}
\begin{split}
M_z\ket{\Psi} =& i\ket{\Psi} \\
\Rightarrow & M_z\mathcal{T}\ket{\Psi}= \mathcal{T}M_z\ket{\Psi}=-i\mathcal{T}\ket{\Psi}. \\
\end{split}
\end{equation}
This means that the STI must have a non-zero mirror Chern number.
Since adding the magnetization term $B_M$ does not break the mirror symmetry and does not close the bulk gap, the mirror Chern number is $C_M = 1$ for every sufficiently small magnetization $M \neq 0$.

This means that time reversal, which maps $M \rightarrow -M$, does not flip the mirror Chern number in this case.
As we show in Sec.~\ref{sec:trandmirrorchernnumber} this is true in general for TCIs with a spinful mirror symmetry.

\section{Layered antiferromagnetic topological insulator}
\label{sec:DTI}

We obtain the Hamiltonian of the layered AFTI, or type-II AFTI, by stacking two-dimensional (2D) Chern-insulating layers with opposite Chern numbers on top of each other.
For the Hamiltonian of the layers we consider two copies of the Qi-Wu-Zhang (QWZ) model \cite{Qi2006, Asbth2016}, which describes a Chern insulator on a square lattice:
\begin{equation}
\begin{split}
&H_\pm(k_x, k_y)=v\left(\sin k_x\sigma_x+\sin k_y \sigma_y\right)\\
&\pm\left(u+t\cos k_x+t\cos k_y\right)\sigma_z
\end{split},
\end{equation}
Here $t$ and $v$ are a real and an imaginary first neighbor hopping, $u$ is an onsite magnetization, $k_x$ and $k_y$ are the crystal momentum components and $\sigma_i$ ($i=x,y,z$) are the Pauli matrices. 
The Hamiltonians of the two layers are related by the time-reversal transformation $\mathcal{T}=i\sigma_y\mathcal{K}$:
\begin{equation}
\mathcal{T}H_-(k_x, k_y)\mathcal{T}^{-1}=H_+(-k_x, -k_y).  
\end{equation}
and thus carry opposite Chern numbers.

The l-AFTI is obtained by stacking the layers on top of each other in the $z$ direction and connecting them via a time-reversal symmetric first neighbor hopping $H_z = i v_z\sigma_y$:
\begin{equation}
\label{eq:dafti}
H_{\rm II}(k_x, k_y, k_z)= \begin{pmatrix}
H_+(k_x, k_y) & H^\pd_z+H_z^\dag e^{-ik_z}\\
H^\dag_z+H_z^\pd e^{ik_z} & H_-(k_x, k_y) 
\end{pmatrix}.
\end{equation}
Throughout the following, we will set $t=v=1$, $u=1.5$, and $v_z=0.1$.

The type-II AFTI Hamiltonian is symmetric under a combination of time-reversal symmetry and a translation by half a unit cell $S(k_z)=\mathcal{T}T_{1/2}(k_z)$:
\begin{equation}
S(k_z)H_{\rm II}(k_x, k_y, k_z)S^{-1}(k_z)=H_{\rm II}(-k_x, -k_y, -k_z),
\end{equation}
where the time-reversal and the translation operator take the form:
\begin{equation}
\mathcal{T}=i \sigma_y \begin{pmatrix}
1 & 0 \\
0 & 1
\end{pmatrix}\mathcal{K}, \quad T_{1/2}(k_z)=\sigma_0\begin{pmatrix}
0 & 1 \\
e^{ik_z} & 0
\end{pmatrix}.
\end{equation}
At $k_z=0$ we have $S^2(k_z=0)=-1$, and as before the system is characterized by a non-trivial $\mathbb{Z}_2$ index, such that it is an AFTI. 
The type A surfaces host topological surface states associated with a Dirac node at $k_z=0$ in the surface band structure, while the type F surfaces are gapped.

The system has the spinless mirror symmetry:
\begin{equation}
M_z(k_z)H_{\rm II}(k_x, k_y, k_z)M_z^{\dag
}(k_z)=H_{\rm II}(k_x, k_y, -k_z),
\end{equation}
where $M_z(k_z)$ is defined as:
\begin{equation}
M_z(k_z)=\sigma_0\begin{pmatrix}
1 & 0 \\
0 & -e^{-ik_z}
\end{pmatrix}.
\end{equation}
In this case, on the $k_z=0$ plane, the Hamiltonian is already block diagonal:
\begin{equation}
H_{\rm II}(k_x, k_y, k_z=0)= \begin{pmatrix}
H_+(k_x, k_y) & 0\\
0 & H_-(k_x, k_y)
\end{pmatrix},
\end{equation}
with the eigenstates of the two blocks $H_\pm(k_x, k_y)$ carrying opposite mirror eigenvalues $m_\pm=\pm 1$. 
The mirror Chern number is then simply given by the difference in Chern number, $C_M=\left[ C(H_+)-C(H_-) \right] /2=1$, between the two layers Hamiltonian. 
Since the mirror Chern number is non-zero this system is also a DTI. 

\section{Invariance of mirror Chern number under time reversal}
\label{sec:trandmirrorchernnumber}

The presence of a non-trivial mirror Chern number has important consequences for the topological properties of the system in the presence of a domain wall, which we model by flipping the magnetization in part of the system.
As mentioned before, we construct the domain wall as an interface between two systems that are related to each other by a time reversal transformation.
If the time-reversal transformation does not change the sign of the mirror Chern number, the 2D domain wall remains gapped except at regions where it intersects the F surfaces.
In that case chiral edge modes occur due to the half-integer Chern number of the type F surfaces.
On the other hand, if the mirror Chern number is flipped by time reversal, the domain wall is an embedded semimetal \cite{Velury2021} with 2D gapless states protected by mirror symmetry.
To understand when the mirror Chern number is invariant under time reversal we can consider a generic TCI with Hamiltonian $H(k_x, k_y, k_z)$ and with mirror symmetry:
\begin{equation}
M_z(k_z)H(k_x, k_y, k_z)M_z^{\dag
}(k_z)=H(k_x, k_y, -k_z).
\end{equation}
On the $k_z=0$ invariant plane, the Hamiltonian can be put in the block-diagonal form:
\begin{equation}
H(k_x, k_y, k_z=0)= \begin{pmatrix}
H_{m_+}(k_x, k_y) & 0\\
0 & H_{m_-}(k_x, k_y)
\end{pmatrix},
\end{equation}
with the two $N\times N$ blocks $H_{m_\pm}(k_x, k_y)$ carrying opposite Chern number $C(H_{m_\pm})= \pm 1$.

In this discussion, we consider a spinful time-reversal operator with $\mathcal{T}^2=-\mathbb{I}$ and we fix the gauge such that the mirror operator commutes with time reversal: $[\mathcal{T}, M_z(k_z=0)]=0$.
With this gauge choice, we say that the mirror symmetry is spinful if the mirror operator has imaginary eigenvalues $m_\pm=\pm i$, whereas the mirror symmetry is spinless if the eigenvalues are real $m_\pm=\pm 1$. Physically a spinful lattice symmetry is used to characterize systems with strong spin-orbit coupling, such as the type-I AFTIs, whereas systems with weak spin-orbit coupling, such as graphene, are characterized using a spinless symmetry. 

We now consider a spinful mirror symmetry with eigenvalues $m_\pm=\pm i$.
In the block-diagonal basis the mirror operator on the mirror plane is:
\begin{equation}
M_z(k_z=0) = i\mathbb{I}_{N\times N}\eta_z,
\end{equation}
where $\eta_z$ is the `which-block' degree of freedom and $\mathbb{I}_{N\times N}$ is the identity acting on the Hilbert space of each block.
We write the time-reversal operator as:
\begin{equation}
\mathcal{T}= i U_{\mathcal{T}}\eta_\alpha\mathcal{K},
\end{equation}
where $\eta_\alpha$ is either the identity or one of the Pauli matrices and $U_{\mathcal{T}}$ is a unitary operator acting on the Hilbert space of each block.
Since we are interested in a spinful time reversal we have the constraint:
\begin{equation}
\mathcal{T}^2= (U^\pd_{\mathcal{T}}U_{\mathcal{T}}^{*})(\eta_{\alpha}^{\pd}\eta_{\alpha}^{*})=-\mathbb{I}_{N\times N}\eta_0,
\end{equation}
which means that either $U^\pd_{\mathcal{T}}U_{\mathcal{T}}^{*}=-\mathbb{I}_{N\times N}$
and $\eta^\pd_\alpha\eta_\alpha^{*}=\eta_0$ or
$U^\pd_{\mathcal{T}}U_{\mathcal{T}}^{*}=\mathbb{I}_{N\times N}$ 
and $\eta^\pd_\alpha\eta_\alpha^{*}=-\eta_0$.  
We now use the fact that time-reversal and mirror symmetry must commute,
\begin{equation}
\mathcal{T}M_z(k_z)\mathcal{T}^{-1}=M_z(-k_z),
\end{equation}
to find $\eta_\alpha$.
On the mirror plane $k_z=0$, we can rewrite the commutation relation as:
\begin{equation}
\label{eq:trcondition}
i(U^\pd_{\mathcal{T}}U_{\mathcal{T}}^{*})(\eta^\pd_\alpha\eta_z\eta_\alpha^{*}) = i\mathbb{I}_{N\times N}\eta_z.
\end{equation}
We have two possible solutions.
If $U^\pd_\mathcal{T}U_{\mathcal{T}}^{*}=-\mathbb{I}_{N\times N}$, Eq.~\eqref{eq:trcondition} takes the form:
\begin{equation}
\label{eq:tretacondition}
\eta_\alpha\eta_z\eta_\alpha^*=-\eta_z.
\end{equation}
In this case the solution is $\eta_\alpha=\eta_x$.
If $U^\pd_\mathcal{T}U_{\mathcal{T}}^{*}=\mathbb{I}_{N\times N}$ then Eq.~\eqref{eq:trcondition}  takes again the form of Eq.~\eqref{eq:tretacondition} but without the minus sign: $\eta_\alpha\eta_z\eta_\alpha^*=\eta_z$. 
Under these conditions the solution is $\eta_\alpha=\eta_y$.

In both cases time reversal exchanges the two blocks of the Hamiltonian, i.e.:
\begin{equation}
\begin{split}
&H^{\rm TR}(k_x, k_y, k_z=0)=\mathcal{T}H(-k_x, -k_y, 0)\mathcal{T}^{-1}= \\
&\begin{pmatrix}
\tau H_{m_-}(-k_x, -k_y)\tau^{-1} & 0\\
0 & \tau H_{m_+}(-k_x, -k_y) \tau^{-1},
\end{pmatrix},
\end{split}
\end{equation}
where $\tau = iU_\mathcal{T}\mathcal{K}$ is the part of the time-reversal operator acting on each single block. 
In general, applying $\tau = iU_\mathcal{T}\mathcal{K}$ flips the Chern number of the two blocks. i.e. $C(\tau H_{m_\pm}\tau^{-1})=-C(H_{m_\pm})$. 
This means that the mirror Chern number does not change sign under time reversal, i.e. $C_M(H^{\rm TR})=C_M(H)$, since the time-reversal transformation is exchanging the two blocks and then flipping their Chern number.

If the mirror symmetry is spinless, the mirror operator has real eigenvalues $m=\pm 1$. 
In the block-diagonal basis, the mirror operator, at $k_z=0$, takes the form:
\begin{equation}
M_z(k_z=0) = \mathbb{I_{N \times N}}\eta_z.
\end{equation}
and, from the commutation relation between the mirror and the time-reversal operator, we obtain:
\begin{equation}\label{eq:second_commutation}
-(U^\pd_{\mathcal{T}}U_{\mathcal{T}}^{*})(\eta^\pd_\alpha\eta^\pd_z\eta_\alpha^{*}) = \mathbb{I}_{N\times N}\eta^\pd_z.
 \end{equation}
In this case if we choose $U^\pd_{\mathcal{T}}U_{\mathcal{T}}^{*}=\mathbb{I}_{N\times N}$, meaning that  $\eta^\pd_\alpha \eta_\alpha^* = -\eta_0 \Rightarrow \eta_\alpha = \eta_y$, Eq.~\eqref{eq:second_commutation} would result in 
$-\eta_y\eta_z\eta^*_y=\eta_z$, which is false. 
The only possibility is then to choose $U^\pd_\mathcal{T}U_\mathcal{T}^{*}=-\mathbb{I}_{N \times N}$, meaning $\eta_\alpha\eta_\alpha^{*}=\eta_0$ and rewrite the commutation condition as:
\begin{equation}
\eta_\alpha\eta_z\eta^*_\alpha=\eta_z.
\end{equation}

The two possible solutions are $\eta_\alpha=\eta_0$ and $\eta_\alpha=\eta_z$ (since $\eta_\alpha\eta_\alpha^{*}=\eta_0$ rules out $\eta_y$).
For both solutions, the time-reversal operator does not exchange the two blocks. 
Now the time-reversed Hamiltonian looks like:
\begin{equation}
\begin{split}
&H^{\rm TR}(k_x, k_y, k_z=0)=\mathcal{T}H(-k_x, -k_y, 0)\mathcal{T}^{-1}=\\
&\begin{pmatrix}
\tau H_{m_+}(-k_x, -k_y)\tau^{-1} & 0\\
0 & \tau H_{m_-}(-k_x, -k_y) \tau^{-1},
\end{pmatrix},
\end{split}
\end{equation}
where the time-reversal transformation $\tau = iU_\mathcal{T}\mathcal{K}$ flips the Chern number of each block. 
This means that, for a TCI with spinless mirror symmetry, time reversal changes the sign of the mirror Chern number, i.e. $C_M(H^{\rm TR})=-C_M(H)$.

\section{Domain walls: construction and topological properties}
\label{sec:domainwall}

We now study the topological properties of domain walls in both types of AFTIs using the tight-binding models of Sec.~\ref{sec:model} and \ref{sec:DTI} and we show numerically that the domain wall is gapped in the type-I AFTI model and an embedded topological semimetal in the type-II AFTI.

A domain wall is created when the magnetization is rotated in part of the system. 
We consider vertical domain walls where the parts of the system with different magnetization are time-reversed partners of each other.
This means that the domain wall respects mirror symmetry.

To construct the domain wall in the type-I AFTI, we consider a ribbon geometry, with conserved momentum $k_y$ and finite size $L_x$ and $L_z$ in the $x$ and $z$ directions. 
We then flip the magnetization at every site $x \ge \frac{L_x}{2}$ by applying time reversal, thus creating a domain wall between the sites $x=\frac{L_x}{2}-1$ and $x=\frac{L_x}{2}$.
The time-reversal transformation changes the sign of the onsite term $B_M \rightarrow -B_M$ in the tight-binding model in  Eq.~\eqref{eq:afti}. 
Since we are only interested in the topological properties of the domain wall, we add periodic boundary conditions (PBC) in the $x$ direction, such that $x=x+L_x$. This creates a new domain wall between $x=L_x-1$ and $x=0$, i.e. across the periodic boundary of the system. 
\begin{figure}
\centering
\includegraphics[width=8.6cm]{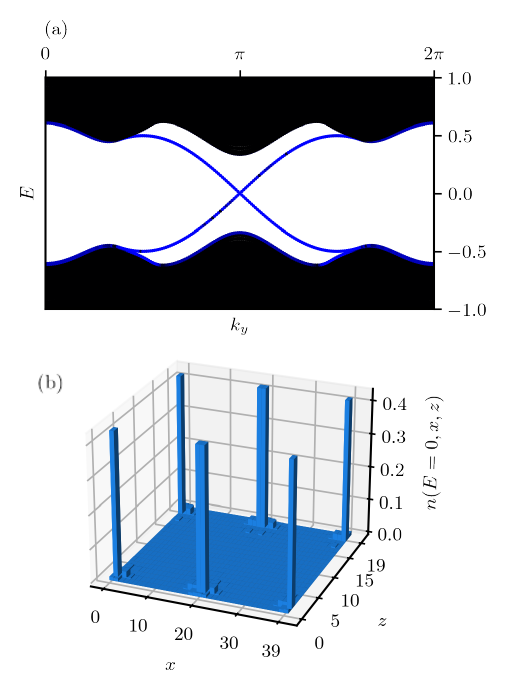}
\caption{
Band structure and LDOS of the type-I AFTI Hamiltonian in Eq.~\eqref{eq:afti} in a ribbon geometry, with conserved momentum $k_y$ and finite size $L_x$ and $L_z$ in the $x$ and $z$ directions, with two domain walls. 
The plots are obtained for $L_x=40$ and $L_z=20$ unit cells, $t=1$, $v=0.5$, $m=2$ and $M=1$. All lengths are in units of the lattice constant $a=1$  and all energies are in units of the nearest neighbor hopping $t=1$.
(a) Band structure of the system. 
The bands crossing the bulk gap (in blue) are associated with gapless states at the intersection between the domain wall and the top/bottom surfaces. 
(b) LDOS of the system at zero energy, showing the localization of the gapless modes, between $x=\frac{L_x}{2}-1$ and $x=\frac{L_x}{2}$ and between $x=0$ and $x=L_x-1$.
}
\label{fig:AFTI}
\end{figure}

We can see the system as two topological insulators connected at the two domain walls, at left- and right-hand side of the defect. 
In this case the mirror symmetry protecting the TCI phase is spinful, which means that the time-reversal transformation does not flip the mirror Chern number of the topological insulator on the right-hand side.
The crystalline topology therefore does not protect any topological state at the domain wall. 
Nonetheless, chiral edge states will appear due to the presence of a half-integer Chern number on the F surfaces \cite{Varnava2021, Rusinov2021}.
If the domain wall were to be moved away from the top and bottom surfaces, for instance if the magnetization is only flipped for sites whose $z$ coordinates are away from the boundaries, then no gapless modes would be formed.

Using \texttt{Kwant} \cite{Groth2014}, a \texttt{Python} library for tight binding calculations, we solve numerically the tight binding model in Eq.~\eqref{eq:afti} in the ribbon geometry. Our code is available on Zenodo \cite{Zenodo_code}.
We find four topological degenerate bands that cross the bulk gap in the band structure of the system, see Fig.~\ref{fig:AFTI}(a). 
By plotting the local density of state (LDOS):
\begin{equation}
n(E, x, z)=\sum_{n, k_y} \delta\left(E-\epsilon_n(k_y)\right)|\braket{x,z}{u_n(k_y)}|^2
\end{equation}
at zero energy, $n(E=0, x, z)$, we see that the band crossings are associated with states localized at the edges of the two domain walls, as shown in Fig.~\ref{fig:AFTI}(b). 

In the type-II AFTI, the presence of dual topology modifies the topological properties of the defect. 
The domain wall hosts 2D gapless topological states protected by the spinless mirror symmetry. 
To show this, we construct the domain wall in a slab geometry with a finite size $L_x$ and two conserved momenta $k_y$ and $k_z$, since we are interested in studying the bulk of the defect. Our code is available on Zenodo \cite{Zenodo_code}.
We flip the magnetization at every site $x \ge \frac{L_x}{2}$ in the model in Eq.~\eqref{eq:dafti} by applying time-reversal symmetry, which exchanges the layer Hamiltonians.  
We add periodic boundary conditions (PBC) in the $x$ direction, thus creating two domain walls, between $x=\frac{L_x}{2}-1$ and $x=\frac{L_x}{2}$ and between $x=0$ and $x=L_x-1$, as before.
\begin{figure}
\centering
\includegraphics[width=8.6cm]{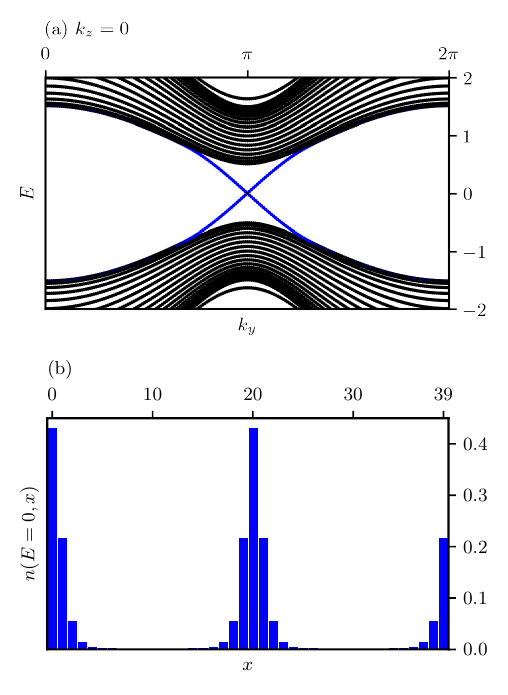}
\caption{
Band structure and LDOS of the type-II AFTI Hamiltonian in Eq.~\eqref{eq:dafti} in a slab geometry, with a finite size $L_x$ and two conserved momenta $k_y$ and $k_z$, with two domain walls. 
The plots where obtained for $L_x=40$, $t=v=1$, $u=1.5$ and $v_z=0.1$. All lengths are in units of the lattice constant $a=1$  and all energies are in units of the nearest neighbor hopping $t=1$.
(a) Band structure of the system on the $k_z=0$ plane. 
The blue bands are associated with gapless states localized at the domain walls. 
(b) LDOS of the system at zero energy which shows that the gapless states are localized at the two domain walls, between $x=\frac{L_x}{2}-1$ and $x=\frac{L_x}{2}$ and between $x=0$ and $x=L_x-1$.
}
\label{fig:DAFTI}
\end{figure}

When the domain wall is introduced in the type-II AFTI model, the two halves of the system, on the left and right sides of the defect at $x=\frac{L_x}{2}$, carry opposite mirror Chern numbers $+1$ and $-1$.
This is due to the fact that the mirror Chern number is flipped by the time-reversal transformation. 
Thus, the domain wall hosts two Dirac cones protected by mirror symmetry.
For the same reason, the second domain wall, formed across the periodic boundary of the system, also hosts two Dirac cones.

The gapless 2D domain wall states are visible in the band structure of Fig.~\ref{fig:DAFTI}(a). 
By plotting the LDOS at zero energy $n(E=0, x)$ we see that these bands are associated with 2D gapless states localized at the domain walls between $x=\frac{L_x}{2}-1$ and $x=\frac{L_x}{2}$ and between $x=0$ and $x=L_x-1$, see Fig~\ref{fig:DAFTI}(b). 
As expected, each domain wall forms a two-dimensional embedded semimetal in the bulk of the system, with gapless states protected by mirror symmetry.

So far we considered only mirror-symmetric domain walls. 
Breaking mirror symmetry will in general gap the topological states at the domain wall. 
Nevertheless, gapless topological states will still be present if the mirror symmetry is locally preserved, e.g. when the domain wall is not a plane but a curved surface. 
For example, consider a spherical domain wall in the bulk of the type-II AFTI, with opposite antiferromagnetic order inside and outside the sphere. 
The equator of the sphere will host gapless edge modes, since the magnetization on opposite sides of the equator are rotated by $\pi$, and the mirror symmetry is locally preserved. 

\section{Conclusions}
\label{sec:conclusion}

In this work we studied the topological properties of domain walls in antiferromagnetic topological insulators.
We considered two different types of AFTIs and we built tight-binding toy models for each of them. 
The first kind (type-I) is similar to MnBi$_{2n}$Te$_{3n+1}$, and is obtained by adding antiferromagnetic order to a strong topological insulator.
The second one (type-II) is made by a stack of weakly-coupled layers with opposite Chern numbers, and it could be realized in van der Waals materials, e.g. in stacks of graphene layers with opposite out-of-plane magnetization \cite{Das2019}.
Both systems are AFTIs with surface states protected by $S$, which is a combination of time-reversal symmetry and translation by half a unit cell. 
Furthermore they are both TCIs with a non-trivial mirror Chern number, on top of a non-trivial $\mathbb{Z}_2$ invariant associated with the antiferromagnetic phase. 
Thus, both systems are dual topological insulators.

However, the two systems differ for the type of mirror symmetry protecting the TCI phase.
The first type of AFTI has a spinful mirror symmetry, while the type-II AFTI has a spinless one. 
As we show in our work, this means that a time-reversal transformation will not change the mirror Chern number of the type-I AFTI, while flipping its sign in the type-II AFTI. 
This has important consequences for the topological properties of domain walls obtained by applying time reversal to part of the system.

In the first type of AFTIs, the bulk of the domain wall is gapped, since time reversal does not flip the mirror Chern number. 
Nonetheless, when the domain wall ends on a F surface, edge states are present at its boundaries, since the F surfaces of the AFTI carry a half-integer Chern number \cite{Varnava2021, Rusinov2021}. 
In a type-II AFTI, the topological properties of the domain wall are modified by the presence of the non-zero mirror Chern number. 
This is due to the fact that time reversal changes sign to the mirror Chern number in TCIs with spinless mirror symmetry.
As a result, a mirror-symmetric domain wall forms an embedded 2D semimetal with gapless topological states protected by mirror symmetry. 
Domain walls which break mirror symmetry, e.g. those with a different orientation or with a magnetization component perpendicular to the domain wall plane, will generically be gapped out.
However, gapless topological states can still be present if the mirror symmetry is locally preserved. 
For example a spherical domain wall in the bulk of the type-II AFTI can host gapless topological states at its equator.
Controlling the geometry of domain walls could thus enable the engineering of the location of topological gapless states within the bulk of these topological magnetic materials.

\section{Acknowledgments}

We acknowledge financial support from the DFG through the W\"urzburg-Dresden Cluster of Excellence on Complexity and Topology in Quantum Matter -- \textit{ct.qmat} (EXC 2147, project-id 390858490). We acknowledge partial support by the Italian Ministry
of Foreign Affairs and International Cooperation, grant
PGR12351 “ULTRAQMAT”. The data that support the findings of this article are openly available \cite{Zenodo_code}, embargo periods may apply.

\bibliography{ref_v2}

\end{document}